\documentclass[aps,prl,reprint,groupedaddress]{revtex4-2}
\usepackage{amsmath,amssymb}
\usepackage{graphicx}

\begin{document}

\title{Variational Gaussian Wave-Packet Dynamics from Constrained Classical Trajectory Bundles}

\author{Taisuke Hasegawa}
\email{Corresponding author: taisukehasegawa@jp-matlantis.com}
\altaffiliation[Present address: ]{Matlantis Corporation, 1-6-1 Otemachi, Tokyo 100-0004, Japan}
\affiliation{Center for Computational Sciences, University of Tsukuba, 1-1-1 Tenno-dai, Tsukuba 305-8577, Japan}

\date{\today}

\begin{abstract}
Variational Gaussian wave-packet dynamics is usually obtained by restricting quantum evolution to Gaussian wave packets. We instead construct finite-$N$ dynamics for a labeled bundle of classical trajectories under a constraint fixing the symplectic-area scale of its covariance. At finite $N$, the bundle itself is the dynamical state: it may be non-Gaussian, and the potential force on each trajectory is evaluated from the full potential without a local harmonic approximation. For sequences whose initial empirical distributions approach a smooth Gaussian density as $N \to \infty$,
the limiting relative motion becomes linear and preserves Gaussianity. With the area scale set to $\hbar/2$, the limiting centroid and width equations coincide with those of the Gaussian time-dependent variational principle, and the limiting phase-space density coincides with the corresponding Gaussian Wigner density for matched initial data. Here $N$ counts trajectories, not orders in a Moyal or $\hbar$ expansion. This phase-space correspondence reveals two complementary constructions of the same Gaussian dynamics: a variational reduction of quantum evolution and the Gaussian large-$N$ limit of a constrained classical trajectory bundle.
\end{abstract}

\maketitle
Variational Gaussian wave-packet dynamics is obtained by applying the time-dependent variational principle (TDVP) to Gaussian wave packets \cite{McLachlan1964,HellerVariational1976,Coalson1990}, yielding centroid and width equations that involve Gaussian averages of the force and Hessian \cite{Coalson1990,Vanicek2023,MoghaddasiFereidani2023}. It differs from Heller's thawed Gaussian approximation, which propagates a Gaussian using a local quadratic expansion of the potential about a classical center \cite{Heller1975,Vanicek2023}. In both approaches, the state remains Gaussian throughout.

Related trajectory-based methods use classical paths to guide moving Gaussian basis functions and locally quadratic Hamiltonians \cite{ShalashilinChild2003}, or to construct cellular and initial-value semiclassical propagators \cite{HellerCellular1991,HermanKluk1984,Antipov2025}. Particularly close are Bohmian approximate-quantum-potential methods, which recast the time-dependent Schr\"odinger equation as trajectory dynamics and approximate the resulting quantum potential or force using the evolving trajectory ensemble \cite{GarashchukRassolov2003AQP,RassolovGarashchuk2004,GarashchukRassolov2019}. These approaches motivate a distinct question: can a finite bundle of classical trajectories itself be the dynamical state, with the Gaussian TDVP emerging as its Gaussian large-$N$ limit?

We begin with $N$ labeled trajectories governed by the same classical Hamiltonian and impose a fixed symplectic-area scale on the covariance of the entire bundle. At finite $N$, the full bundle---rather than a Gaussian wave-packet ansatz \cite{PattanayakSchieve1994} or a closed cumulant hierarchy \cite{Shigeta2006}---is the dynamical state. The fixed-area condition constrains the covariance but not the remaining bundle statistics, so finite bundles may be non-Gaussian. It acts collectively on the bundle covariance rather than imposing coordinate or velocity constraints familiar from classical mechanics and constrained molecular dynamics \cite{Goldstein1980,Ryckaert1977,Andersen1983}, and forces determined by the instantaneous bundle preserve it. No wave function or explicit quantum potential is introduced in the finite-$N$ formulation: the potential force on each trajectory is evaluated from the full potential, without a local harmonic approximation.

For sequences whose initial empirical phase-space distributions approach a smooth Gaussian density as $N\to\infty$, the nonlinear residual force no longer appears in the limiting relative motion, which becomes linear and preserves Gaussianity. When the symplectic-area scale is set to $\hbar/2$, the limiting centroid and width equations coincide with the Gaussian TDVP, and the full limiting phase-space density coincides with the corresponding Gaussian Wigner density for matched initial data. Numerical tests on a 20-dimensional coupled Morse model illustrate the finite-$N$ dynamical effect of the constraint and compare selected observables with the derived Gaussian large-$N$ limit.

\noindent\emph{Finite trajectory bundle and fixed-area condition.---}
We consider a bundle of $N$ labeled classical trajectories, each governed by the same Hamiltonian in mass-weighted coordinates,
\begin{equation}
	h(q_i,\pi_i) =  \frac{1}{2}\pi_i^{\mathrm T}\pi_i + V(q_i),
	\label{eq:hamiltonian}
\end{equation}
with $q_i,\pi_i \in \mathbb{R}^d$, $i=1,\ldots,N$. We denote the bundle average by $\langle a\rangle=N^{-1}\sum_i a_i$. For each trajectory, we separate centroid and relative variables by writing $q_i=q_c+x_i$, $\pi_i=\pi_c+p_i$, where $q_c=\langle q\rangle$ and $\pi_c=\langle \pi\rangle$. By construction, $\langle x\rangle=\langle p\rangle=0$. Trajectory indices are suppressed inside bundle averages; for example, $\langle xx^{\mathrm T}\rangle=N^{-1}\sum_i x_i x_i^{\mathrm T}$. The centered covariance blocks of the bundle are
\begin{equation}
	X=\langle xx^{\mathrm T}\rangle,\qquad C=\langle px^{\mathrm T}\rangle,\qquad P=\langle pp^{\mathrm T}\rangle.
	\label{eq:widths}
\end{equation}
With this convention, $\langle xp^{\mathrm T}\rangle=C^{\mathrm T}$. 
The full centered phase-space covariance matrix and the standard symplectic matrix are
\begin{equation}
	\Sigma = \begin{pmatrix} X & C^{\mathrm T} \\ C & P \end{pmatrix},\qquad
	J = \begin{pmatrix} 0 & I \\ -I & 0 \end{pmatrix},
	\label{eq:sigmaJ}
\end{equation}
where $I$ is the $d \times d$ identity matrix.
We restrict attention to nondegenerate bundles for which $\Sigma$ is positive definite. Motivated by Williamson's symplectic normal form for positive-definite matrices \cite{Williamson1936,Arvind1995}, we impose the following fixed symplectic-area condition,
\begin{equation}
	\Sigma J\Sigma = \kappa J,\quad \kappa > 0.
	\label{eq:fixedarea}
\end{equation}
The quantity $\sqrt \kappa$ sets the common prescribed symplectic-area scale of the covariance. Since $\Sigma$ is positive definite, $X$ is also positive definite. We define
$A=CX^{-1}$ and introduce the residual momentum $r_i=p_i-Ax_i$. Its covariance is
$\Pi=\langle rr^{\mathrm T}\rangle=P-CX^{-1}C^{\mathrm T}$.
By construction, $\langle rx^{\mathrm T}\rangle=0$; thus, $r_i$ is the component of
the relative momentum remaining after subtracting the component linearly correlated with the displacement. 
In these variables, the fixed-area condition takes the simple form
\begin{equation}
	\Sigma J\Sigma = \kappa J \quad \Longleftrightarrow \quad
	A=A^{\mathrm T},\qquad \Pi=\kappa X^{-1},
	\label{eq:equivalence}
\end{equation}
as shown in Appendix A.
Introducing $G \equiv \Pi-\kappa X^{-1}$, we write the fixed-area conditions as
\begin{equation}
	G =0,\qquad A=A^{\mathrm T}.
	\label{eq:conditions}
\end{equation}
These are covariance-level relations; they neither determine the remaining bundle statistics nor impose a Gaussian form. The fixed-area condition is imposed on the bundle as a whole, while each trajectory remains labeled and classical. 

\noindent\emph{Constraint forces.---}
We now derive constraint forces that preserve the fixed-area conditions during the finite-bundle evolution. Let $f_i$ denote the centered potential force, excluding the constraint force,
\begin{equation}
	f_i=-\nabla V(q_c+x_i)+\langle \nabla V(q_c+x)\rangle.
	\label{eq:centeredforce}
\end{equation}
By construction, $\langle f\rangle=0$. At each time, we decompose $f_i$ into a component that is linear in the displacement within the bundle and a residual component,
\begin{equation}
	f_i=M_N x_i+u_i,\qquad M_N=\langle fx^{\mathrm T}\rangle X^{-1},
	\label{eq:force_decomposition}
\end{equation}
By construction, $\langle u\rangle=0$ and $\langle ux^{\mathrm T}\rangle=0$. The term $M_Nx_i$ is the linear force component determined by the force–displacement correlation of the current bundle. The term $u_i$ is the residual force not represented by this linear component within the bundle.
For compactness, we define
\begin{equation}
	R=\langle ur^{\mathrm T}\rangle.
	\label{eq:Rdef}
\end{equation}
The matrix $R$ measures the correlation between the residual force and the residual momentum. As shown below, the symmetric part of $R$ gives the direct contribution of the residual force to the change of $G$. 

Because $G=0$ is quadratic in the momenta, we do not use it directly as a virtual-work constraint. Instead, we begin with the preservation condition $\dot G =0$. On the fixed-area conditions $G=0$ and $A=A^{\mathrm{T}}$, the time derivative obtained in the absence of constraint forces is
\begin{equation}
\label{eq:appendix-result}
	\left.\dot G\right|_{f_{\mathrm{con}}=0}=R+R^{\mathrm T}.
\end{equation}
Thus the direct contribution of the residual force vanishes when $R_+ = R + R^{\mathrm{T}}=0$. Because $r=p-Ax$
and $\langle ux^{\mathrm T} \rangle=0$, one has
$R=\langle ur^{\mathrm T} \rangle =\langle up^{\mathrm T} \rangle $. Thus, at fixed positions, $R_+=0$ is linear in the momenta and can be used in the d’Alembert construction.
The virtual-work construction in Appendix B gives the associated constraint force
\begin{equation}
	f_{+,i}=\Lambda_{+}u_i,\qquad \Lambda_{+}^{\mathrm T}=\Lambda_{+},
	\label{eq:fplus}
\end{equation}
where $\Lambda_{+}(t)$ is a symmetric multiplier.
Including this force, the time derivative of $G$ becomes
\begin{equation}
	\left.\dot G\right|_{f_{\mathrm{con}}=f_{+}}=R_+ + 
	\Lambda_{+}R+R^{\mathrm T}\Lambda_{+}.
	\label{eq:Gdot_fplus}
\end{equation}
The condition $R_+=0$ removes the direct contribution of the residual force to the change of $G$. 
Define $R_- \equiv R-R^{\mathrm T}$. On $R_+=R+R^{\mathrm T}=0$, the remaining contribution 
in Eq.~\eqref{eq:Gdot_fplus} can be written as
\begin{equation}
    \Lambda_{+}R+R^{\mathrm T}\Lambda_{+} = \frac{1}{2} \left[ \Lambda_{+},R_{-} \right].
	\label{eq:commutator}
\end{equation}
The symmetric multiplier $\Lambda_{+}(t)$ must satisfy the consistency condition
$\dot R_+=0$ and generally depends on the instantaneous bundle. Requiring preservation of $G=0$ without assuming a special commutation relation therefore singles out $R_-=0$
as the remaining compatibility condition. It is not an additional prescribed width constraint.
The d’Alembert constraint force associated with $ R_-=0$ is
\begin{equation}
	f_{-,i}=\Lambda_{-}u_i.
	\label{eq:fminus}
\end{equation}
Here $\Lambda_{-}$ is an antisymmetric multiplier, $\Lambda_{-}^{\mathrm T}=-\Lambda_{-}$.
Together, $ R_+=0 $ and $ R_-=0 $ give $R=0$, and the full multiplier
$\Lambda = \Lambda_+ + \Lambda_-$ must satisfy $\dot R=0$.
With $R=0$, neither the residual potential force nor the combined constraint force, $f_\Lambda$, changes $G$. Hence, with $ f_{\Lambda} = f_+ + f_- $, $\dot G = 0$.
The two constraint-force components combine as
$f_{\Lambda,i}=\Lambda u_i$, $\Lambda=\Lambda_{+}+\Lambda_{-}$, 
where $\Lambda(t)$ is a general matrix.
The remaining fixed-area condition is $A=A^{\mathrm T}$.
A separate virtual-work construction gives the constraint force
\begin{equation}
	f_{\Omega,i}=\Omega X^{-1}x_i,\qquad \Omega^{\mathrm T}=-\Omega.
	\label{eq:fOmega}
\end{equation}
The antisymmetric multiplier $\Omega (t)$ is determined by requiring preservation of this condition,
$ \dot{A}-\dot{A}^{\mathrm T}=0$.
On $G=0$ and $A=A^\mathrm{T}$, the force $f_\Omega$ does not contribute to $\dot G$. 
Therefore, with $f_{\mathrm{con},i}=f_{\Lambda,i} + f_{\Omega,i}$,
\begin{equation}
	\dot G =0.
	\label{eq:Gdot_constraint_omega}
\end{equation}
Because $\langle u\rangle=0$ and $\langle x\rangle=0$, the bundle average of the constraint force vanishes: $\langle f_{\mathrm{con}}\rangle=0$. Moreover, using $R=0$, $A=A^{\mathrm T}$, and $\Omega^{\mathrm T}=-\Omega$, one obtains
\begin{equation}
	\langle \pi^{\mathrm T}f_{\mathrm{con}}\rangle=\langle p^{\mathrm T}f_{\mathrm{con}}\rangle=0.
	\label{eq:zerowork}
\end{equation}
Thus the instantaneous bundle-averaged power of the constraint force vanishes. The power on an individual trajectory, however, need not vanish. For a quadratic potential, the centered potential force is exactly linear in the displacement, and therefore $u_i = 0$; the resulting linear classical evolution also preserves the fixed-area condition, giving $\Omega = 0$. The constraint force therefore vanishes and the constrained dynamics can differ from the unconstrained dynamics only through the anharmonic part of the potential.

With these constraint forces, each member of the finite bundle obeys
\begin{equation}
	\dot q_i=\pi_i,\qquad \dot\pi_i=-\nabla V(q_i)+\Lambda u_i+\Omega X^{-1}x_i.
	\label{eq:full_eom}
\end{equation}
Because both constraint-force terms have zero bundle average, the equations separate into centroid and relative motion. The centroid equations are
\begin{equation}
	\dot q_c=\pi_c,\qquad \dot\pi_c=-\langle \nabla V(q_c+x)\rangle.
	\label{eq:centroid}
\end{equation}
The relative variables obey
\begin{equation}
	\dot x_i=p_i,\qquad \dot p_i=M_Nx_i+(I+\Lambda)u_i+\Omega X^{-1}x_i.
	\label{eq:relative_eom}
\end{equation}
With $\Lambda$ and $\Omega$ required to satisfy the consistency conditions $\dot R=0$ and $\dot A - \dot A^\mathrm{T}=0$,
respectively, Eqs.~\eqref{eq:full_eom}--\eqref{eq:relative_eom} define the finite-$N$ constrained dynamics of the labeled trajectory bundle.
For initial data satisfying the fixed-area relation and the associated compatibility condition, this dynamics preserves the prescribed symplectic area. The trajectories remain classical but are coupled through collective forces determined by the instantaneous bundle. 

\noindent\emph{Calibration to the quantum area scale.---}
Up to this point, $\sqrt \kappa$ has specified the prescribed symplectic-area scale of the covariance in a finite-bundle dynamics built from classical trajectories. We now set $\kappa=\hbar^2/4$,
so that this area scale is $ \sqrt \kappa = \hbar/2$.
This choice does not change the classical form of the trajectory-level equations: each bundle member remains a constrained classical trajectory. The resulting covariance relation has the same algebraic form as that for the Wigner transform of a pure Gaussian wave packet \cite{Ohsawa2017,Arvind1995} and is consistent with saturation of the Robertson--Schr\"odinger uncertainty relation \cite{Robertson1929,Schrodinger1930}.
This choice calibrates only the covariance scale; it does not impose a Gaussian form on a finite bundle.
With this calibration, the finite-bundle equations describe constrained classical trajectory dynamics whose covariance carries the quantum symplectic-area scale $\hbar/2$. We next examine the Gaussian large-$N$ limit of these calibrated finite-bundle equations.

\noindent\emph{Gaussian large-$N$ limit.---}
We now consider a sequence of finite constrained trajectory bundles whose initial empirical phase-space distributions converge to a smooth, nondegenerate Gaussian density as $N \to \infty$. In taking this limit, the empirical averages appearing below are assumed to converge to the corresponding Gaussian averages. We first derive the limiting equations for a Gaussian density and then show that the resulting evolution preserves Gaussianity. In this section, trajectory labels are suppressed: $x$, $p$, and $r$ denote relative phase-space variables distributed according to the limiting density, while $\langle \cdot \rangle_G$ denotes averaging over that density.

We now establish the principal result in this Gaussian large-$N$ limit. The derivation below shows that the force--displacement matrix becomes the negative Gaussian-averaged Hessian, the nonlinear residual force is removed from the relative motion by the limiting constraint, and the resulting linear flow preserves Gaussianity. After setting $\kappa=\hbar^2/4$, the limiting density is the Wigner density generated by the Gaussian TDVP for matched initial data.

For a Gaussian limiting density, $r=p-Ax$ is a linear transformation of $(x,p)$, so $x$ and $r$ are jointly Gaussian. Together with $\langle rx^\mathrm{T} \rangle_G=0$, this makes them independent. Gaussian integration by parts then gives
\begin{equation}
	M_N \to M_G=-\langle \nabla^2 V(q_c+x)\rangle_G,
	\label{eq:MG}
\end{equation}
which is symmetric. Let $u_G=f_G-M_Gx$ denote the corresponding residual force, where $f_G$ is the centered force in the Gaussian limit.
Since $u_{G}$ depends only on $x$, and since $x$ and $r$ are independent with $\langle r \rangle_G = 0$, one has
$ R_G \equiv \langle u_{G}r^\mathrm{T} \rangle_G = 0$.
Requiring this relation to be preserved gives the limiting condition for the multiplier
$\Lambda_{G}$. As shown in Appendix C, the consistency condition gives
\begin{equation}
	U_G(I+\Lambda_G)^\mathrm{T}=0, \quad U_G=\langle u_G u_G^\mathrm{T} \rangle_G,
	\label{eq:contG}
\end{equation}
or equivalently $ (I+\Lambda_G)U_G = 0$.
Thus the residual force does not appear in the relative equation of motion in the Gaussian limit. This conclusion does not require $U_G$ to be nonsingular.
Since $M_G=M_G^\mathrm{T}$, preservation of $A=A^\mathrm{T}$ gives 
$ \Omega_G X^{-1} + X^{-1} \Omega_G = 0$. Because $X$ is positive definite and
$\Omega_G^{\mathrm{T}} = -\Omega_G$, this implies $\Omega_G = 0$.
The limiting relative motion therefore obeys
\begin{equation}
	\dot x = p, \quad \dot p = M_Gx.
	\label{eq:Grel}
\end{equation}
Although $M_G(t)$ may depend on time through the evolving density, it is the same matrix for every phase-space point. Equation~\eqref{eq:Grel} therefore defines a linear time-dependent relative flow, which maps an initially Gaussian density to a Gaussian density; the centroid equations add only a common phase-space translation.

Using $C=AX$, $A=A^\mathrm{T}$, and $\Pi = \kappa X^{-1}$, the width equations become
\begin{equation}
	\dot X = AX + XA, \quad \dot A = M_G - A^2 + \kappa X^{-2}.
	\label{eq:width_eom}
\end{equation}
With the complex symmetric width matrix
\begin{equation}
	Z=A+i\sqrt{\kappa}\,X^{-1},
	\label{eq:Zdef}
\end{equation}
these equations combine into
\begin{equation}
	\dot Z=M_G-Z^2.
	\label{eq:Zdot}
\end{equation}
The centroid equations are
\begin{equation}
	\dot q_c=\pi_c,\qquad \dot\pi_c=-\langle \nabla V(q_c+x)\rangle_G.
	\label{eq:gaussian_centroid}
\end{equation}
After the calibration $\kappa = \hbar^2/4$, the collective limiting equations,
 Eqs.~\eqref{eq:Zdot} and \eqref{eq:gaussian_centroid}, coincide with the width and centroid equations of variational Gaussian wave-packet dynamics, equivalently with those obtained from the Gaussian time-dependent variational principle (TDVP)
 \cite{McLachlan1964,HellerVariational1976,Coalson1990,Vanicek2023,MoghaddasiFereidani2023}. Because the limiting density remains Gaussian and is completely determined by its centroid and covariance, this correspondence extends to the full phase-space density: for matched initial data, the limiting trajectory density coincides with the Wigner density of the corresponding variational Gaussian wave packet. Within this setting, this identifies the Gaussian TDVP as the Gaussian large-$N$ limit of the constrained classical-bundle equations. At finite $N$, the dynamics consists instead of a finite set of labeled constrained trajectories.

\noindent\emph{Numerical tests.---}
We use a 20-dimensional coupled Morse-oscillator model~\cite{MoghaddasiFereidani2023} to test two complementary aspects of the theory: the finite-$N$ dynamical effect of the fixed-area constraint and the finite-$N$ behavior relative to the Gaussian large-$N$ limit derived above. All calculations use $\kappa=\hbar^2/4$, and all plotted quantities are expressed in natural units (n.u.).
The finite-bundle equations are integrated using only force evaluations along the trajectories; the multiplier equations are solved iteratively at each time step without evaluating Hessians. For each $N$, ten initial bundles are generated independently from the same target Gaussian phase-space density and then adjusted to reproduce its centroid and covariance to numerical precision. 
Each initial bundle also satisfies $A=A^{\mathrm T}$, $\Pi=\kappa X^{-1}$, and $R=0$.
The bundles are then propagated under the constrained classical equations without enforcing a Gaussian phase-space distribution. Details of the model, time-integration algorithm, and preparation of the initial bundles are given in the Supplemental Material~\cite{SupplementalMaterial}.
At fixed $N=2^{16}$, we isolate the effect of the constraint by propagating each initial bundle twice from the same initial trajectory coordinates and momenta, once with and once without the constraint force.

\begin{figure}
	\includegraphics[width=0.95\columnwidth]{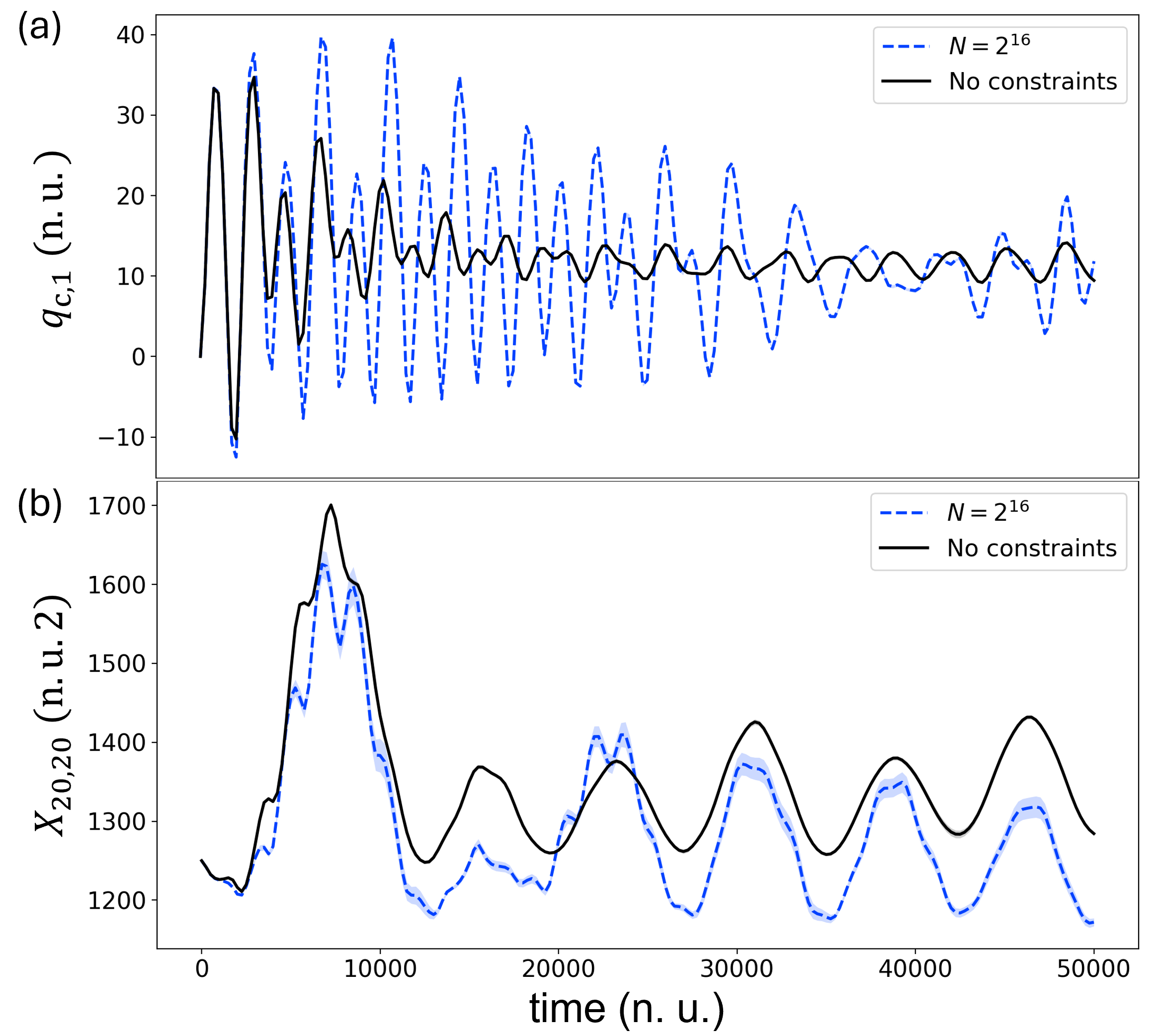}
	\caption{ 
		Effect of the fixed-area constraint at $N=2^{16}$.
		Each of ten initial bundles is propagated with and without the constraint from identical initial trajectory coordinates and momenta. Lines show realization averages, and shaded bands show one standard deviation. (a) Centroid coordinate $q_{c,1}(t)$. (b) Position-covariance element ${X_{20,20}(t)}$.}
	\label{fig:m1}
\end{figure}

Figures~\ref{fig:m1}(a) and \ref{fig:m1}(b) show the realization-averaged centroid coordinate $q_{c,1}(t)$ and position-covariance element $X_{20,20}(t)$, respectively.
In the unconstrained calculations, the centroid-oscillation amplitude decreases markedly over the plotted interval.
The constrained $q_{c,1}(t)$ curve retains a larger oscillation amplitude, while the constrained and unconstrained $X_{20,20}(t)$ curves also evolve differently.
Across the ten paired realizations, the fixed-area constraint therefore produces systematic changes in both plotted observables.

\begin{figure}
	\includegraphics[width=0.95\columnwidth]{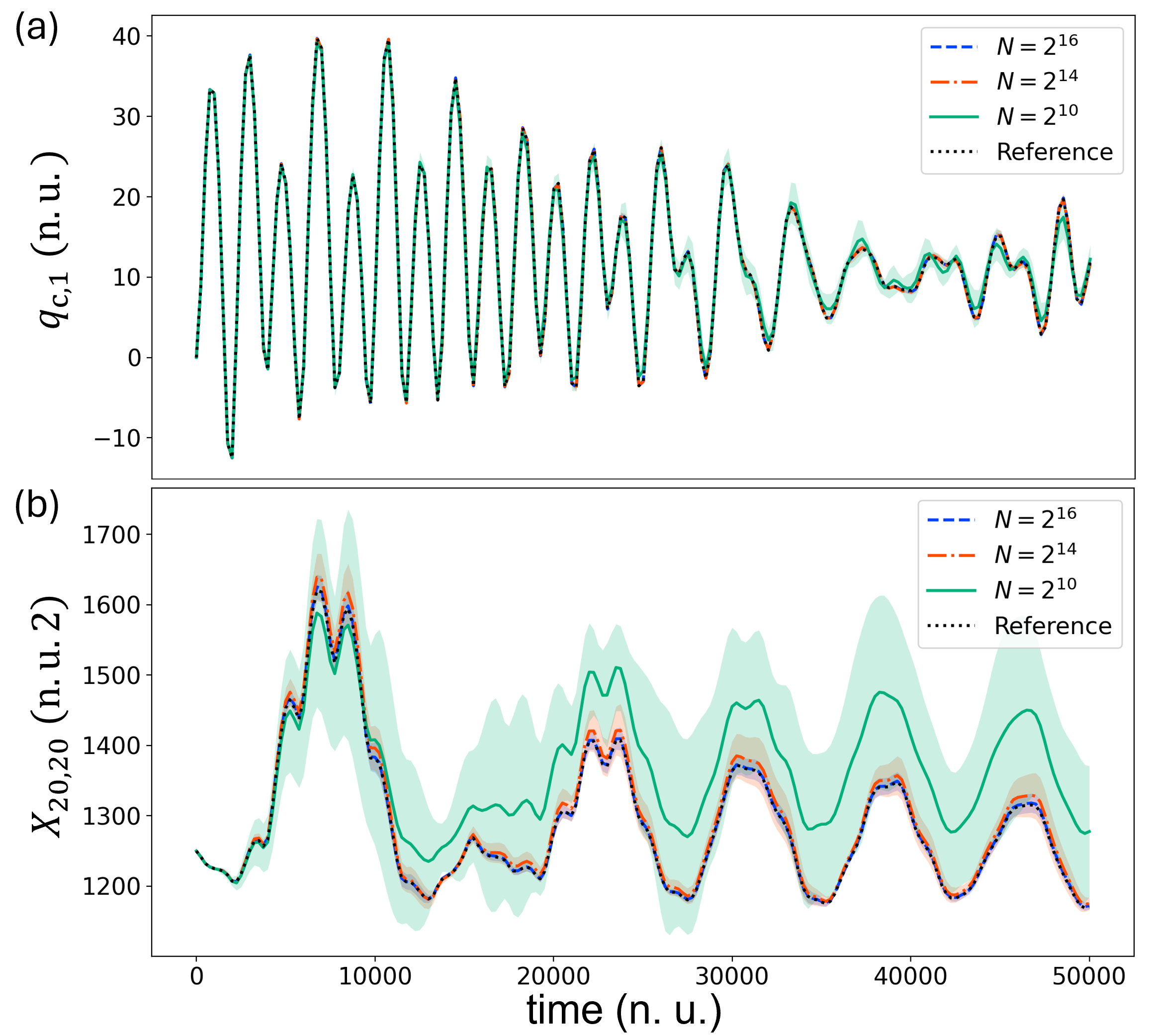}
	\caption{Finite-$N$ dependence of the constrained dynamics relative to the Gaussian large-$N$ limit. Lines and shaded bands show realization averages and one standard deviation over ten initial bundles. The Gaussian TDVP curve representing the analytically derived Gaussian large-$N$ limit is shown for comparison. (a) Centroid coordinate $q_{c,1}(t)$. (b) Position-covariance element ${X_{20,20}(t)}$.}
	\label{fig:m2}
\end{figure}

We next compare constrained bundles with $N=2^{10}$, $N=2^{14}$, and $N=2^{16}$ against this Gaussian limit, represented by a reference curve obtained from the Gaussian TDVP equations~\cite{MoghaddasiFereidani2023}. At $\kappa = \hbar^2/4$, these equations are equivalent to the limiting collective equations, Eqs.~\eqref{eq:width_eom} and \eqref{eq:gaussian_centroid}. 
Figures~\ref{fig:m2}(a) and \ref{fig:m2}(b) show the realization-averaged $q_{c,1}(t)$ and $X_{20,20}(t)$, respectively.
Over the tested range, the realization-averaged $q_{c,1}(t)$ curves depend only weakly on $N$.
For the plotted $X_{20,20}(t)$ curves, a finite-$N$ trend is apparent: the realization-averaged $N=2^{10}$ curve departs from the reference at later times, whereas the $N=2^{14}$ and $N=2^{16}$ curves lie closer to it, with the $N=2^{16}$ curve remaining close throughout the plotted interval.

Together, the constrained--unconstrained comparison shows a systematic dynamical effect of the constraint at finite $N$. The results across $N$ for the two plotted observables are consistent with the Gaussian large-$N$ limit.
The full phase-space-density correspondence in the Gaussian large-$N$ limit is established analytically; the numerical calculations examine finite-$N$ behavior relative to that limit in these selected observables. Step-size tests reported in the Supplemental Material~\cite{SupplementalMaterial} confirm stable time integration and numerical preservation of the prescribed symplectic area.

\noindent\emph{Discussion.---}
The principal conceptual implication is that the Gaussian TDVP admits a complementary trajectory-bundle interpretation. Within the stated Gaussian large-$N$ setting, the same phase-space dynamics that follows from a variational restriction of quantum evolution also arises from a constrained classical trajectory bundle. Because the correspondence holds for the full Gaussian Wigner density rather than only for the centroid and width equations, the associated pure Gaussian wave function is fixed up to the global phase absent from the Wigner representation \cite{Ohsawa2017,Kubo1964}. The correspondence applies to the Gaussian limit, not to the finite-$N$ bundle itself, which remains a distinct and potentially non-Gaussian dynamical object.

The distinction from earlier trajectory-based constructions concerns both the starting point and where the additional structure is introduced. Bohmian approximate-quantum-potential methods start from the time-dependent Schr\"odinger equation and approximate the quantum potential or force that appears in its hydrodynamic formulation \cite{GarashchukRassolov2003AQP,RassolovGarashchuk2004,GarashchukRassolov2019}. Local-quadratic methods attach wave-packet structure to classical trajectories \cite{Heller1975,ShalashilinChild2003}, whereas cellular and initial-value methods attach propagator structure \cite{HellerCellular1991,HermanKluk1984,Antipov2025}. The present theory begins with $N$ trajectories governed by a classical Hamiltonian and constrains the covariance of the bundle itself; no wave function or explicit quantum potential enters the finite-$N$ equations. The potential force on each trajectory is evaluated from the full potential, while the collective constraint forces preserve the fixed-area condition, have zero bundle average and zero bundle-averaged power, and vanish for quadratic potentials. The Gaussian-averaged Hessian and linear relative motion emerge only in the Gaussian large-$N$ limit. Because the fixed-area condition does not close the finite-$N$ distribution, the bundle retains higher-order statistics not fixed by its covariance.

The independent roles of $N$ and $\sqrt{\kappa}$ sharpen this distinction. The former specifies the number of trajectories, whereas the latter sets the prescribed symplectic-area scale. Unlike the Moyal phase-space expansion in $\hbar$ \cite{Groenewold1946,Moyal1949}, varying $N$ does not change an expansion order; $\hbar$ enters only through the calibration $\sqrt{\kappa}=\hbar/2$. In one degree of freedom, the condition becomes $XP-C^2=\kappa$ and, at $\kappa=\hbar^2/4$, recovers the constant-uncertainty condition introduced in CUMD \cite{Hasegawa2016}, which was later extended to three-dimensional condensed-phase systems \cite{Hasegawa2023}. The numerical calculations demonstrate a dynamical effect of the constraint and behavior of selected observables consistent with the derived Gaussian large-$N$ limit, but they do not establish general convergence of the finite-$N$ dynamics. Nor are finite-$N$ departures identified here as systematic corrections to exact quantum dynamics. Rather, the construction provides a common framework for examining when finite, potentially non-Gaussian bundle dynamics agrees with or departs from the Gaussian TDVP in anharmonic multidimensional systems.

\begin{acknowledgments}
OpenAI ChatGPT (GPT-5.6 and GPT-5.5) was used as a supportive tool for conceptual brainstorming, language translation, code generation, and organizing the scientific exposition. The author provided detailed instructions, reviewed all suggested changes, and verified the final text against the analytical derivations and numerical implementation. The author retains full responsibility for all content and conclusions of this manuscript.
\end{acknowledgments}

\bibliography{export_v2.2_prl_format_clean_reference_checked}

\bigskip
\appendix
\section*{APPENDIX}

\setcounter{equation}{0}
\renewcommand{\theequation}{A\arabic{equation}}
\noindent \emph{Appendix A: Fixed-area condition in residual-momentum variables.---}
From the definitions $A=CX^{-1}$ and $\Pi=P-CX^{-1}C^\mathrm{T}$,
one has $C=AX$ and $P=AXA^\mathrm{T}+\Pi$, together with $\langle rx^\mathrm{T} \rangle =0$.
The covariance matrix therefore factorizes as
\begin{equation}
	\label{eq:appendix-result1}
	\Sigma=\begin{pmatrix} I & 0 \\ A & I \end{pmatrix}
	\begin{pmatrix} X & 0 \\ 0 & \Pi \end{pmatrix}
	\begin{pmatrix} I & A^{\mathrm T} \\ 0 & I \end{pmatrix}.
\end{equation}
Substitution into $\Sigma J\Sigma=\kappa J$ gives, from the $(1,1)$ and $(2,1)$ blocks,
\begin{equation}
\label{eq:appendix-result2}
X(A-A^{\mathrm T})X=0,\qquad \Pi X=\kappa I.
\end{equation}
Because $X$ is positive definite, these relations imply $A=A^{\mathrm T}$, $\Pi=\kappa X^{-1}$. Conversely, substituting $A=A^{\mathrm T}$ and $\Pi=\kappa X^{-1}$ into the factorized form of $\Sigma$ gives $\Sigma J\Sigma=\kappa J$. Thus the fixed-area condition is equivalent to
\begin{equation}
 A=A^{\mathrm T},\quad G\equiv\Pi-\kappa X^{-1}=0.
\end{equation}

\par\medskip
\setcounter{equation}{0}
\renewcommand{\theequation}{B\arabic{equation}}
\noindent \emph{Appendix B: Virtual-work construction of the constraint forces.---}
We give the virtual-work construction of the forces associated with the compatibility condition $R=0$ and
the fixed-area condition $A=A^\mathrm{T}$.
The centered potential force is decomposed as
$ f_i=M_Nx_i+u_i$, where $\langle u \rangle =0$, 
$\langle ux^\mathrm{T} \rangle =0$, and 
$R=\langle ur^\mathrm{T} \rangle$.
On $G=0$ and $A=A^{\mathrm T}$, the centered potential force gives the direct contribution
\begin{equation}
\left.\dot G\right|_{f_{\mathrm{con}}=0}=R+R^{\mathrm T}.
\end{equation}
The direct residual-force contribution therefore vanishes when
$ R_+ \equiv R+R^\mathrm{T}=0$.
At fixed positions, $\delta r_i=\delta p_i-\delta Ax_i$.
Since $\langle ux^\mathrm{T} \rangle =0$, the term involving 
$\delta A$ drops out, and therefore 
$\delta R= \langle u \delta p^\mathrm{T} \rangle$.
Pairing $\delta R_{+}=\delta R+\delta R^{\mathrm T}$ with a symmetric multiplier $\Lambda_{+}^{\mathrm T}=\Lambda_{+}$ gives
\begin{equation}
	\label{eq:appendix-result3}
	\delta W_{+}=\frac{1}{2}\operatorname{tr}\Lambda_{+}\delta R_{+}=\langle \delta p^{\mathrm T}\Lambda_{+}u\rangle.
\end{equation}
The corresponding constraint force is therefore $f_{+,i}=\Lambda_{+}u_i$.
 The condition $R_{+}=0$ alone does not yet guarantee preservation of $G=0$, because the associated constraint force also contributes to $\dot G$.
 On $R_{+}=0$, the contribution of $f_{+}$ is 
\begin{equation}
\label{eq:appendix-result4}
(\dot G)_{f_+}=
\Lambda_{+}R+R^{\mathrm{T}} \Lambda_{+} = \frac{1}{2}
[ \Lambda_{+}, R_{-} ],
\end{equation}
where $R_- \equiv R-R^\mathrm{T}$.
The multiplier $\Lambda_+(t)$ must satisfy the consistency condition $\dot R_+=0$ and generally depends on the instantaneous bundle. For nonzero $R_-$, vanishing of the commutator in Eq.~\eqref{eq:appendix-result4} would require a special commutation relation between $\Lambda_+$ and $R_-$. Requiring preservation without assuming such a relation therefore identifies $R_-=0$ as the remaining compatibility condition.
Pairing $\delta R_-$ with an antisymmetric multiplier
 $\Lambda_-^\mathrm{T} = - \Lambda_-$ gives
\begin{equation}
	\delta W_-=\frac{1}{2} \mathrm{tr}(\Lambda_-\delta R_-)
	=\langle \delta p^\mathrm{T} \Lambda_- u \rangle.
\end{equation}
The associated force is $f_{-,i}=\Lambda_-u_i$.
Together, $R_+=0$ and $R_-=0$ give $R=0$. The two forces combine as $f_{\Lambda,i} = \Lambda u_i$ and $\Lambda = \Lambda_+ + \Lambda_-$.
On $R =0$, $\langle f_\Lambda r^\mathrm{T} \rangle 
= \Lambda R = 0$,
and its transpose also vanishes. Hence $f_\Lambda$ gives no
additional contribution to $ \dot G$.
For the remaining fixed-area condition, the variation of 
$A=CX^{-1}$ at fixed position is $\delta A = 
\langle \delta p x^\mathrm{T} \rangle X^{-1}$.
Pairing the antisymmetric part $A-A^\mathrm{T}$ with an 
antisymmetric multiplier $\Omega^\mathrm{T} = - \Omega$ gives
\begin{equation}
	\delta W_\Omega=-\frac{1}{2}
	 \mathrm{tr}[\Omega \delta (A-A^\mathrm{T})]
	=\langle \delta p^\mathrm{T} \Omega X^{-1}x \rangle.
\end{equation}
The associated force is $f_\Omega = \Omega X^{-1}x$.
Under this force, $(\dot A)_\Omega = 
\langle f_\Omega x^\mathrm{T} \rangle X^{-1} = \Omega X^{-1}$,
and therefore $(\dot r)_\Omega = f_\Omega - (\dot A)_\Omega x =0$.
Thus $f_\Omega $ does not change $\Pi$ or $G$.
Both components of the total constraint force have zero bundle
average. Moreover,
$\langle p^\mathrm{T} f_\Lambda \rangle
= \mathrm{tr}(\Lambda R)=0$, and
$\langle p^\mathrm{T} f_\Omega \rangle
= \mathrm{tr}(\Omega A^\mathrm{T})=0$.
Since $\langle f_{\mathrm{con}} \rangle = 0 $, it follows that
$\langle \pi^\mathrm{T} f_{\mathrm{con}} \rangle =  
\langle p^\mathrm{T} f_{\mathrm{con}} \rangle = 0$.

\par\medskip
\setcounter{equation}{0}
\renewcommand{\theequation}{C\arabic{equation}}
\noindent \emph{Appendix C: Algebra used in the Gaussian large-$N$ limit.---}
We collect the algebraic steps used in the Gaussian large-$N$ limit. Consider a sequence of finite bundles whose initial empirical phase-space distributions converge to a smooth, nondegenerate Gaussian density as $N \to \infty$, and assume that the empirical averages below converge to the corresponding Gaussian averages. We first derive the limiting equations for a Gaussian density; the linear relative dynamics obtained below then shows that Gaussianity is preserved. Averages over the limiting density are denoted by $\langle \cdot \rangle_G$, and trajectory labels are suppressed below.
For a Gaussian limiting density, Gaussian integration by parts gives
\begin{equation}
	\langle f x^\mathrm{T} \rangle_G = 
	-\langle \nabla^2 V(q_c + x)\rangle_G X.
\end{equation}
Thus, in this limit, the finite-bundle linear-force matrix becomes
\begin{equation}
\label{eq:appendix-result5}
M_N=\langle fx^{\mathrm T}\rangle X^{-1}\longrightarrow M_G=-\langle \nabla^2 V(q_c+x)\rangle_G.
\end{equation}
 In this limit, $x$ and $r$ are independent because they are jointly Gaussian and $\langle rx^{\mathrm T}\rangle_G=0$. Since $u_G$ depends only on $x$ and $\langle r\rangle_G=0$, one has $R_G \equiv \langle u_Gr^{\mathrm T}\rangle_G=0$. In the finite-$N$ dynamics, $\Lambda$ must satisfy the consistency condition $\dot R=0$. In this Gaussian limit, the preservation condition becomes
\begin{equation}
\label{eq:appendix-result6}
(I+\Lambda_G)U_G=0,\qquad U_G=\langle u_G u_G^{\mathrm T}\rangle_G.
\end{equation}
When $U_G$ is nonsingular, this implies $I+\Lambda_G=0$, and hence $(I+\Lambda_G)u_G=0$. If $U_G$ is rank deficient, the multiplier $\Lambda_G$ is fixed only in the directions in which the residual force fluctuates. In directions where the residual force has no fluctuation, there is no component to cancel, and the remaining freedom in $\Lambda_G$ does not change the force $(I+\Lambda_G)u_G$. Thus $(I+\Lambda_G)u_G=0$ even when $U_G$ is rank deficient.

For the $\Omega_G$ component of the constraint force, preservation of $A=A^{\mathrm T}$ gives
\begin{equation}
\label{eq:appendix-result7}
\Omega_G X^{-1}+X^{-1}\Omega_G=0.
\end{equation}
Together with $\Omega_G^{\mathrm T}=-\Omega_G$ and $X>0$, this gives $\Omega_G=0$. The relative motion is therefore governed by the linear force,
\begin{equation}
\label{eq:appendix-result8}
\dot x=p,\qquad \dot p=M_Gx.
\end{equation}
Although $M_G(t)$ may depend on time through the evolving density, it is common to all phase-space points; the resulting linear time-dependent flow preserves an initially Gaussian density.

Using the limiting relative equations, $\dot x=p$ and $\dot p=M_Gx$, the position-covariance derivative is
\begin{equation}
	\dot X
	=
	\left\langle \dot x x^{\mathrm T}
	+x\dot x^{\mathrm T}\right\rangle_G
	=
	C+C^{\mathrm T}
	=
	AX+XA.
	\label{eq:appendix-result9}
\end{equation}
The momentum--position covariance $C=\langle p x^\mathrm{T} \rangle_G$ satisfies
\begin{equation}
	\dot C
	=
	\left\langle \dot p x^{\mathrm T}
	+p\dot x^{\mathrm T}\right\rangle_G
	=
	M_GX+P.
	\label{eq:appendix-Cdot}
\end{equation}
Since $A=CX^{-1}$, its derivative is
\begin{equation}
	\begin{aligned}
		\dot A
		&=
		\dot C X^{-1}
		-A\dot X X^{-1} \\
		&=
		M_G+PX^{-1}
		-A(AX+XA)X^{-1} \\
		&=
		M_G-A^2+\kappa X^{-2},
	\end{aligned}
	\label{eq:appendix-result10}
\end{equation}
where we used
$P=AXA^{\mathrm T}+\Pi=AXA+\kappa X^{-1}$.
Equations~\eqref{eq:appendix-result9} and
\eqref{eq:appendix-result10} are the width equations given in
Eq.~\eqref{eq:width_eom} of the main text.

With the complex symmetric width matrix
\begin{equation}
	Z=A+i\sqrt{\kappa}\,X^{-1},
	\label{eq:appendix-result11}
\end{equation}
the inverse position covariance evolves according to
\begin{equation}
	\frac{d}{dt}X^{-1}
	=
	-X^{-1}\dot X X^{-1}
	=
	-\left(X^{-1}A+AX^{-1}\right).
	\label{eq:appendix-Xinv-dot}
\end{equation}
It follows that
\begin{equation}
	\begin{aligned}
		\dot Z
		&=
		M_G-A^2+\kappa X^{-2}
		-i\sqrt{\kappa}
		\left(X^{-1}A+AX^{-1}\right) \\
		&=
		M_G-
		\left(A+i\sqrt{\kappa}\,X^{-1}\right)^2 \\
		&=
		M_G-Z^2.
	\end{aligned}
	\label{eq:appendix-result12}
\end{equation}
Together with the centroid equations
\begin{equation}
	\dot q_c=\pi_c,\qquad
	\dot\pi_c=-\left\langle
	\nabla V(q_c+x)\right\rangle_G,
	\label{eq:appendix-result13}
\end{equation}
this gives the phase-space form of variational Gaussian
wave-packet dynamics, equivalently the Gaussian TDVP dynamics,
after the calibration $\kappa=\hbar^2/4$.

\end{document}